\documentclass[aps,12pt,final,notitlepage,oneside,onecolumn,nobibnotes,nofootinbib,superscriptaddress,noshowpacs,centertags]{revtex4}

\usepackage{amsmath,amssymb,epsfig,placeins,subfigure,wrapfig,indentfirst,makeidx}
\usepackage[cp1251]{inputenc}
\usepackage[english]{babel}
\usepackage{amsfonts,amssymb}
\usepackage{mathrsfs}
\usepackage{dsfont}

\begin{document}

\title{Spectrum for the electric dipole \\
 which nonradially falling into a black hole}

\author{Alexander Shatskiy\footnote{shatskiy@asc.rssi.ru}}
\affiliation{P.N. Lebedev Physical Institute, Astro Space Center,
84/32 Profsoyuznaya st., Moscow, GSP-7, 117997, Russia}

\author{I.D. Novikov}
\affiliation{P.N. Lebedev Physical Institute, Astro Space Center, 84/32 Profsoyuznaya st., Moscow, GSP-7, 117997, Russia}
\affiliation{The Niels Bohr International Academy, The Niels Bohr Institute, Blegdamsvej 17, DK-2100 Copenhagen, Denmark}

\author{Alexandr Malinovsky}
\affiliation{P.N. Lebedev Physical Institute, Astro Space Center, 84/32 Profsoyuznaya st., Moscow, GSP-7, 117997, Russia}

\date{\today}

\begin{abstract}

The electromagnetic bremsstrahlung spectrum of the dipole which falls along a spiral orbit into the Schwarzschild black hole was found.
The characteristic features of this electromagnetic spectrum can be used for determination of the black hole mass.
This new way (if implemented) provides higher accuracy of the determination of the black hole mass.
Also these features in the spectrum can be used for the determination of the some physical processes in the black hole magnetosphere
and in the accretion disk.
It is also shown that the asymptotic behavior of this spectrum (at high frequencies) is practically independent
from the impact parameter of the falling dipole.

\end{abstract}

\maketitle

\section{Introduction}
\label{s1}

As it is well known (see eg~\cite{Thorn1998, Frolov1998}) black holes no "hair".
Therefore, all electromagnetic fields from multipole moments will disappear if the system of charges approaches to the black hole horizon.
For the point charge at rest the electric field was considered by Linet~\cite{Linet1976}.
It has been shown, in particular, that the field of a point charge approaches to the field of the charged black hole
(with the same charge) when the charge approaches to the horizon.
Hence it follows that all fields of all electric and magnetic multipole moments should be radiated when the charge
(or a system of charges or currents) approaches to the black hole horizon.

For accelerated motion of monopole (single charge) the energy loss is determined mainly by bremsstrahlung.
This radiation is a dipolar, since the radiated field components are inversely proportional to the $c^2$ (square of the speed of light).

The existence of dipole radiation is not obvious for a massive dipole falls into a black hole
(due to the fact that both of the charges of the dipole are moving and accelerating in the same direction, and the signs of the charges are opposite).
However, the increasing of the space-time curvature leads to the emission of the dipole type of radiation (see section~\ref{s4}).

Earlier, a similar radiation was investigated in many papers --
see for example~\cite{Zerilli1970, Davis1971, Ross1971, Teukolsky1972, Yakovlev1975, Dymnikova1977, Martel2008};
We do this a slightly different method, which is much more convenient for numerical calculations -- see~\cite{Shatskiy2013, Shatskiy2014}. 

In the paper~\cite{Nail2014} it was found the bremsstrahlung radiation for a charge passing (with a constant velocity) through the wormhole.

In this paper we will solve our problem in general case of non-radial dipole which free falling into a Schwarzschild black hole.

\section{Law of motion of a free-falling particle}
\label{s2}

Schwarzschild metric for a nonrotating and uncharged black hole has the form:
\begin{eqnarray}
ds^2=\left(1-\frac{r_g}{r}\right) c^2\, dt^2 - \left(1-\frac{r_g}{r}\right)^{-1}\, dr^2 - r^2 (d\theta^2 + \sin^2\theta\, d\varphi^2) .
\label{metrkn}\end{eqnarray}
Here: ${r_g=2GM/c^2}$ -- the radius of the black hole horizon, $G$ -- gravitational constant, $M$ -- the mass of the black hole.

Let us remind the law of motion for the test particles in the Schwarzschild gravitational field.
We suppose that the particle moves in the equatorial plane ${(\theta =\pi /2)}$.
Let us write the geodesic equation for the particle (see.~\cite{Brill1960, Landau1988}, \S 87):
\begin{eqnarray}
\frac{du_i}{ds}=\frac{1}{2}\frac{\partial g_{kl}}{\partial x^i} u^k u^l .
\label{GEO}\end{eqnarray}
Hence for the metric (\ref{metrkn}) and ${i = 0}$ (corresponding to the time coordinate ${ct}$) we have the integral of motion\footnote{In our work, the Latin indices run a series of spatial coordinates and time coordinates, and Greek indices run a series of spatial coordinates.}:
${u_0 := \epsilon =const}$ (the specific energy the particles);
and for $i$, which corresponds coordinates $\varphi$ we have the integral of motion: ${u_\varphi = L/(mc) := h \epsilon =const}$
(here $L$ -- angular momentum, $m$ -- mass and $h$ -- the impact parameter of the particle).

We assume that the fall of the particles happens from the initial radius $r_0$ and the result of this fall is the capture of the particles by a black hole.
We also assume that the mass of the particle is much smaller than the mass of the black hole: ${m<<M}$.
From equations (\ref{GEO}) and taking into account the identity ${u_i u^i \equiv 1}$, we have:
\begin{eqnarray}
\frac{c dt}{ds}\equiv u^0 = \epsilon /(1-r_g/r) \label{u_t} , \\
\frac{dr}{ds}\equiv u^r = -\sqrt{\epsilon^2 - (1-r_g/r)(1+h^2\epsilon^2/r^2)} \label{u_r} , \\
\frac{d\varphi}{ds}\equiv u^\varphi = -h\epsilon /r^2 \label{u_varphi} .
\end{eqnarray}
Here ${r(t)}$ -- the current radial coordinate of the particle.

To find the highest possible value\footnote{This corresponds to the minimum-possible value of the impact parameter of the particle,
in which it still will not be captured by a black hole.} of the impact parameter of the particle ${h_{max}}$
(in which it will be captured by a black hole), we proceed similarly the work~\cite{Novikov1967}.

The value of ${h_{max}}$ is defined as the root of the equation ${[u^r]^2_{(h,r)}=0}$,
where this root at the same time should be a point of minimum for function ${[u^r]^2_{(r)}}$.
Thus the point of minimum $r_m$ determined by solution of the equation:
\begin{eqnarray}
(\epsilon^2 - 1)r_m^2 + (2 - 1.5\epsilon^2)r_m r_g - r_g^2 = 0 .
\label{h_max0}\end{eqnarray}
From the two roots of this equation we need to choose a smaller
(which corresponding to the plus sign and the minimum distance to the black hole).
Then the value of ${h_ {max}}$ is determined by the expression:
\begin{eqnarray}
h_{max}^2 = \frac{(\epsilon^2 r_m - r_m + r_g)r_m^2}{\epsilon^2 (r_m - r_g)} .
\label{h_max1}\end{eqnarray}
At ${\epsilon = 1}$ (particle is at rest at infinity) we have:
\begin{eqnarray}
h_{max}^{(\epsilon = 1)} = 2 r_g .
\label{h_max_e1p}\end{eqnarray}
In the limit ${\epsilon\to\infty}$ (for photons), we have:
\begin{eqnarray}
h_{max}^{(\epsilon\to\infty)} = \frac{3\sqrt{3}r_g}{2} .
\label{h_max_f}\end{eqnarray}

The minimum valid value for $\epsilon$ is determined by the inequality: ${\epsilon^2>8/9}$.

\section{Radiation of a charge}
\label{s3}

For the covariant 4-vector potential $A_i$ (for the electromagnetic field), we have two invariants: ${inv_1 = A_i u^i}$ and ${inv_2 = A_i A^i}$.
In accordance with the gauge invariance of the 4-vector potential we choose the calibration ${\tilde A_i}$
in the comoving (free-falling) reference system so that\footnote{In general, the gauge transformation ${A_i \to A_i +\partial_i f(x^k)}$
can not be made to vanish by all three spatial components ${A_\gamma}$, but in the comoving (for incident particle)
reference system it is possible to do -- see for example~\cite{Landau1988}, \S 65.} ${\tilde A_\gamma := 0}$.
Thus, in the comoving reference frame we have: ${\tilde A_0 = q/R}$,
where ${q}$ -- the particle charge, ${R=|\vec r_{ob} - \vec r|}$, ${\vec r_{ob}}$ and ${\vec r}$
-- the radius-vectors of the observer and of the charge (respectively).
Hence, according to the Lorentz transformations, we obtain for the covariant spatial components of the potential
in the Schwarzschild reference frame near the particle:
\begin{eqnarray}
A_0 = \frac{q}{R} u_0\, ,\quad A_\gamma = \frac{q}{R} u_\gamma .
\label{inv_A} \end{eqnarray}
To get the covariant components of the electric field ${F_{0\ gamma}\equiv\partial_0 A_\gamma - \partial_\gamma A_0}$ 
it is necessary to differentiate the expressions (\ref{inv_A}) with respect to the time coordinate ${ct_{ob}}$
(at the observation point) and with respect to the spatial coordinates of the radius vector ${\vec r_{ob}}$ (at the observation point).
Because $u_0$ and $u_\varphi$ are constants (see section~\ref{s2}) and ${u_\theta =0}$, we have:
\begin{eqnarray}
F_{0r} = \frac{q}{R} \partial_{0} u_r .
\label{F_ij} \end{eqnarray}
In the expression (\ref{F_ij}) we have discarded a member with the asymptotic ${\propto 1/R^2}$
and allowed only a member with the required asymptotic behavior: ${\propto 1/R}$
in the field of the electromagnetic wave\footnote{Therefore, to find the radiated field components we need to differentiate only ${u_i}$.} (emw):
\begin{eqnarray}
F^{\rm emw}_{0\gamma}F_{\rm emw}^{0\gamma} \propto \frac{1}{R^2} \, , \quad F^{\rm emw}_{\alpha\gamma}F_{\rm emw}^{\alpha\gamma} \propto \frac{1}{R^2} .
\label{ass_F_ti} \end{eqnarray}
The asymptotic behavior ${\propto 1/R}$ in the electromagnetic field corresponds to the field of the wave propagating
along the vector at infinity ${\vec R}$.
In this case, the field asymptotic behavior ${\propto 1/R^2}$ does not correspond to electromagnetic waves, becouse it does not satisfy
the conservation of the energy flux through a sphere with the radius ${r>>r_g}$.
At the same time, at distances which is much larger than the size of the radiating system (ie maximum radial coordinates of the particle),
wave approximation works and we can talk about radiation of photons by the particle.
We are interested not only the processes at large distances in the wave zone, but (mostly) the processes near the black hole and near the particle.
When you change locations of the charge (or a system of the charges) in
Schwarzschild coordinates the corresponding change of electromagnetic field propagates at the speed of light.
We are interested in the change in electromagnetic field, which corresponds to the asymptotic behavior of the wave field
${(\propto 1/R)}$ and then we will talk about such a field -- as the radiation field of of the electromagnetic wave with wave vector ${k^i}$.
At the same time, we remember that we are considering processes occuring not only in the wave zone.

To find the Fourier transform of the radiation field we proceed similarly to\footnote{Here we take into account that up to time
${t_{ob} = 0}$ the particle at rest (for a stationary observer), therefore the emitted fields are absent.
Since any function can be represented as a superposition of symmetric and antisymmetric parts,
then we extrapolating the function ${F_{t\gamma}}$ to the negative half-time even and odd way --
in the half-sum we get zero, and each of the parts (even and odd) we use for the Fourier transform.}
our work~\cite{Shatskiy2013}:
\begin{eqnarray}
F^{(w)}_{0\gamma} = \frac{1}{2}\int\limits_0^\infty \left[\sin\left(wt_{ob}\right)+\cos\left(wt_{ob}\right)\right]\, F_{0\gamma} \, dt_{ob} .
\label{F_w_ti_2}\end{eqnarray}
Here, the value ${w\equiv c k_0}$ has the physical meaning of covariance, time component of the wave 4-vector\footnote{The values
of the contravariant components of the wave 4-vector $k^i$ (corresponding to the electromagnetic wave which propagating in the plane of the equator) are obtained from expressions (\ref{u_t}-\ref{u_varphi}) by the replacement ${u^i\to\epsilon k^i}$ in the limit
${\epsilon\to\infty}$.}, ${k_0}$ -- integral of the motion for the photon (analogue $u_0$ from (\ref{GEO})).

Because the observer with the radial coordinate ${r_{ob}}$ and 4-velocity ${U^i}$ (relative to predetermined reference system)
registers the frequency of the electromagnetic field
${\hat w:= c k_i U^i}$, then in general the frequency of which registers the fixed observer is equal to
${\hat w = w/\sqrt{1-r_g/r_{ob}}}$.
From the point of view of the stationary observer (at radius $r_{ob}$) the formula (\ref{F_w_ti_2}) can be rewritten as:
\begin{eqnarray}
F^{({\hat w})}_{0\gamma} = \frac{1}{\sqrt{2}}\int\limits_0^\infty
\cos\left({\hat w}t_{ob}\sqrt{1-\frac{r_g}{r_{ob}}} - \frac{\pi}{4}\right) \, F_{0\gamma} \, dt_{ob} .
\label{F_w_ti_2-2}\end{eqnarray}
At ${r_{ob}>>r_g}$ we have ${\hat w \approx w}$.

In formulas (\ref{F_w_ti_2}) and (\ref{F_w_ti_2-2}) the point of time $t_{ob}$ is the time of arrival of the wave to the observer.
At the same time $t_{ob}$ should be carried differentiation (to obtain the field ${F_{0\gamma}}$ from potential $A_\gamma$).
Therefore, in expression (\ref{F_w_ti_2-2}) we can use integration by parts\footnote{In formula
(\ref{F_w_ti_2-3}) it was taken into account that on the horizon for the dipole we have ${A_{\gamma}^{r=r_g}=0}$ -- see.~(\ref{inv_A_d1}).}:
\begin{eqnarray}
F^{({\hat w})}_{0\gamma} =  -\frac{A_{\gamma}^{(t_{ob}=0)}}{2} + \frac{{\hat w}\sqrt{1-r_g/r_{ob}}}{\sqrt{2}}\int\limits_0^\infty
\sin\left({\hat w}t_{ob}\sqrt{1-\frac{r_g}{r_{ob}}} - \frac{\pi}{4}\right) \, A_{\gamma}^{(t_{ob})} \, dt_{ob} .
\label{F_w_ti_2-3}\end{eqnarray}
In order to determine the time $t_{ob}$ we consider two consecutives near events in the Schwarzschild reference system: \\
1) the field measurement by observer at the time of arrival of the particle at the point with the radial coordinate $r_1$; \\
2) the field measurement by observer at the time of arrival of the particle at the point with the radial coordinate $r_2$.
And ${r_1-r_2:=dr>0}$.

The corresponding total changing of the time ${dt_{ob}}$ consists of two components:

1) Period of time\footnote{Hereafter the subscript "${{}_ f}$"\, corresponds to the retarded potential of the field.}
${\Delta t_{f}}$, which corresponds to the difference in length of paths ${\delta l_1}$ and ${\delta l_2}$
for distribution (with speed of light) a retarded potential of the radiated field.
Moreover, the retarded potential extends from the points with radial coordinates $r_1$ and $r_2$
(respectively) to the observer (with the radial coordinate ${r_{ob}}$).
If you know the impact parameter for the photon $h_{f}$, then from expressions (\ref{u_t}) and (\ref{u_r})
at ${\epsilon\to\infty}$ (for distribution of retarded potential) we get by clock a stationary observer at infinity
${\Delta t_{f} = \sqrt{g_{00}^{ob}}(u^0/u^r) dr/c}$:
\begin{eqnarray}
\Delta t_{f} = \frac{-\sqrt{1-r_g/r_{ob}}\, dr}{c(1-r_g/r)\sqrt{1 - (1-r_g/r)h_{f}^2/r^2}} .
\label{dt_ph}\end{eqnarray}

2) the time required to move the particle from the point $r_1$ to the point $r_2$ (by clock a stationary observer)
-- see (\ref{u_t}) and (\ref{u_varphi})).

Thus, the sum of these two components gives:
\begin{eqnarray}
dt_{ob} = \frac{-\sqrt{1-r_g/r_{ob}}\, dr}{c(1-r_g/r)\sqrt{1 - (1-r_g/r)h_{f}^2/r^2}}
+ \frac{-\sqrt{1-r_g/r_{ob}}\, dr}{c(1-r_g/r)\sqrt{1 - (1-r_g/r)(1/\epsilon^2 + h^2/r^2)}} .\quad
\label{dt_ob}\end{eqnarray}

\section{The radiation of the dipole}
\label{s4}

The field of the retarded potentials of the dipole is determined by the standard way:
as a superposition of the fields from each of the charges of the dipole.
The dipole moment $d_0$ is also defined by the standard way:
\begin{eqnarray}
d_0 := q\cdot\delta l = const ,
\label{d_0} \end{eqnarray}
here ${\delta l}$ -- constant length of the rigid dipole which is defined in the framework of general relativity.
To determine the ${\delta l}$ we assume that in the reference frame which associated with the dipole one of its charges emits a gamma quantum,
and the other charge through the proper time ${\delta\tau}$ this gamma quantum are absorbed.
Then the square of the interval ${ds^2}$ between these two events gets (in the comoving and Schwarzschild reference systems, respectively):
\begin{eqnarray}
ds^2 = c^2\delta\tau - \delta l^2 = 0 ,   \quad      ds^2 = g_{ij} \delta x^i \delta x^j = 0 .
\label{d_1} \end{eqnarray}
From here to determine the magnitude ${\delta l}$ we use another invariant -- see~(\ref{GEO}):
\begin{eqnarray}
c \delta\tau = \delta x^i u_i = inv
\label{d_2} \end{eqnarray}
We consider three possible variants for the dipole orientation in space:
along ${\delta r}$, along ${\delta\theta}$ and along ${\delta\varphi}$.
For the dipole orientation along the increment ${\delta x^\alpha}$ (in spatial coordinates) we introduce the notation:
${\delta x^\alpha := \delta l \cdot f^\alpha}$.
From equations (\ref{d_1}) and (\ref{d_2}) we obtain for these three possible orientations of the dipole:
\begin{eqnarray}
f^r = \frac{(1-r_g/r)}{\epsilon - u^r} , \quad
f^\theta = \frac{\sqrt{1-r_g/r}}{\epsilon r} , \quad
f^\varphi = \frac{\sqrt{1-r_g/r}}{\epsilon\left(r + h\sqrt{1-r_g/r}\right)} .
\label{d_3} \end{eqnarray}
Then, in the formula (\ref{inv_A}), which written for the dipole, it will be a corresponding increment
${\delta u_\gamma}$ for 4-velocity under the parallel translation ${u_\gamma}$ along ${\delta x^\alpha}$ in a curved space-time:
\begin{eqnarray}
A_\gamma = \frac{q}{R}\,\delta u_\gamma \equiv \frac{q}{R} \left( \partial_\alpha u_\gamma - \Gamma^k_{\gamma\alpha} u_k \right) \delta x^\alpha =
\frac{d_0 f^\alpha}{R} \left( \partial_\alpha u_\gamma - \Gamma^k_{\gamma\alpha} u_k \right)
\label{inv_A_d0}\end{eqnarray}
Here ${\Gamma^i_{kl} := \frac{1}{2}g^{im}\left(\partial_l g_{mk} + \partial_k g_{ml} - \partial_m g_{kl}\right)}$ -- Christoffel symbols.

The Fourier transform of the electric dipole field at infinity we obtain by substituting (\ref{inv_A_d0}) in the formula (\ref{F_w_ti_2-3}).

The retarded potentials (\ref{inv_A}) and (\ref{inv_A_d0}) in the leading order are proportional to ${1/c}$ (inversely proportional the speed of light),
because in the nonrelativistic limit ${(\epsilon\to 1 , \,\, r_g/r\to 0)}$.
The spatial components for 4-velocity inversely proportional to the speed of light ${(u^\gamma\propto {\rm v}/c)}$,
and zero (time) component of the particle 4-velocity in the main approximation by $c$ does not depend.
Therefore the emitted fields of the charge and the dipole in the main approximation are corresponded to the emission of the dipole type radiation
(which inversely proportional to the square of the speed of light).

\section{The calculation of the spectral density of the dipole radiation}
\label{s5}

We pass to dimensionless units: ${r_g=1}$ and ${c=1}$. Let us
consider the simplest (in terms of computations) dipole
orientation\footnote{Here we have designated:
${\delta^\alpha_\theta}$ -- Kronecker symbol.}: ${\delta x^\alpha
= \delta^\alpha_\theta\cdot\delta l\cdot f^\theta}$. In this case,
the orientation of the dipole in spherical coordinates stored over
time. Let the observer has the coordinates: ${r_{ob} >> r_g}$,
${\theta_{ob} =\pi /2}$, ${\varphi_{ob} =0}$ (at the equator
plane). Then from (\ref{inv_A_d0}) we obtain:
\begin{eqnarray}
A_\gamma = -\delta_\gamma^\theta \cdot\frac{d_0 f^\theta r u^r}{R} = \frac{\delta_\gamma^\theta \, d_0}{R}\,
\sqrt{\left(1-\frac{r_g}{r}\right)-\left(1-\frac{r_g}{r}\right)^2\left(\frac{1}{\epsilon^2}+\frac{h^2}{r^2}\right)} .
\label{inv_A_d1}\end{eqnarray}
Note that the form of the expression (\ref{inv_A_d1}) ensures the convergence of the integral (\ref{F_w_ti_2-3})
throughout the range of the radial coordinate $r$
and time ${t_{ob}}$ (despite the divergence of expressions for ${dt_{ob}}$ near the horizon of the black hole).

For brevity, we will talk about the photons (instead of the above-mentioned fields of retarded potential of the electromagnetic wave),
thus all characteristics of the photon is determined from the black hole (in the wave zone).
Accordingly, for a dipole we introduce the quantities subscripts "${{}_d}$", and for the field of the retarded potential -- "${{}_f}$".

Denoting: ${\xi :=1/r}$, from (\ref{u_t}-\ref{u_varphi}) we obtain for the total deflection angle of the photon:
\begin{eqnarray}
\varphi_{f} = \int\limits^{\xi_d}_{\xi_{ob}} \frac{h_{f}\, d\xi}{\sqrt{1-h_{f}^2\xi^2(1-\xi)}} + \varphi_d
\label{varphi1} \end{eqnarray}
During the movement the particle (dipole) emits photons in opposite directions,
but we are interested in only those photons which are reach the observer,
ie at position ${\varphi_{f}=\varphi_{ob}:=0}$ and ${\xi_{ob}=1/r_{ob}}$.
In this case, the initial coordinates for photons are: ${r_d}$, ${\varphi_d(r_d)}$ and time ${t_d(r_d)}$ (for each of the photons -- its value $r_d$).
To calculate the trajectory of these photons, we also need to know their impact parameters ${h_{f}(r_d)}$.
Therefore, to calculate the change of the impact parameter ${dh_{f}}$ (between different photons -- during the motion of a particle)
we obtain from (\ref{varphi1}) the differential relation:
\begin{eqnarray}
d\varphi_{f}(h_f, \xi_d) = \frac{\partial\varphi_{f}}{\partial\xi_d}\, d\xi_d + \frac{\partial\varphi_{f}}{\partial h_{f}}\, dh_{f} + d\varphi_d(\xi_d) = 0 ,
\label{varphi2}\end{eqnarray}
where
\begin{eqnarray}
\frac{\partial\varphi_{f}}{\partial\xi_d} = \frac{h_{f}}{\sqrt{1-h_{f}^2\xi_d^2(1-\xi_d)}} , \label{varphi_fd}\\
\frac{\partial\varphi_{f}}{\partial h_{f}} = \int\limits^{\xi_d}_{\xi_{ob}} \frac{d\xi}{\left[1-h_{f}^2\xi^2(1-\xi)\right]^{3/2}} ,
\label{varphi_ff} \end{eqnarray}
and taking into account (\ref{u_r}) and (\ref{u_varphi}) we have:
\begin{eqnarray}
d\varphi_d = \frac{-h_{d}\, d\xi_d}{\sqrt{1-(\epsilon^{-2}+h_{d}^2\xi_d^2)(1-\xi_d)}} .
\label{varphi3} \end{eqnarray}
At ${\varphi_d =0}$ we have ${h_{f}=0}$ (the initial moment ${t_d=0}$).
Thus, from the relation (\ref{varphi2}) we can calculate ${h_{f}}$.

\begin{figure}
\includegraphics[width=0.98\textwidth]{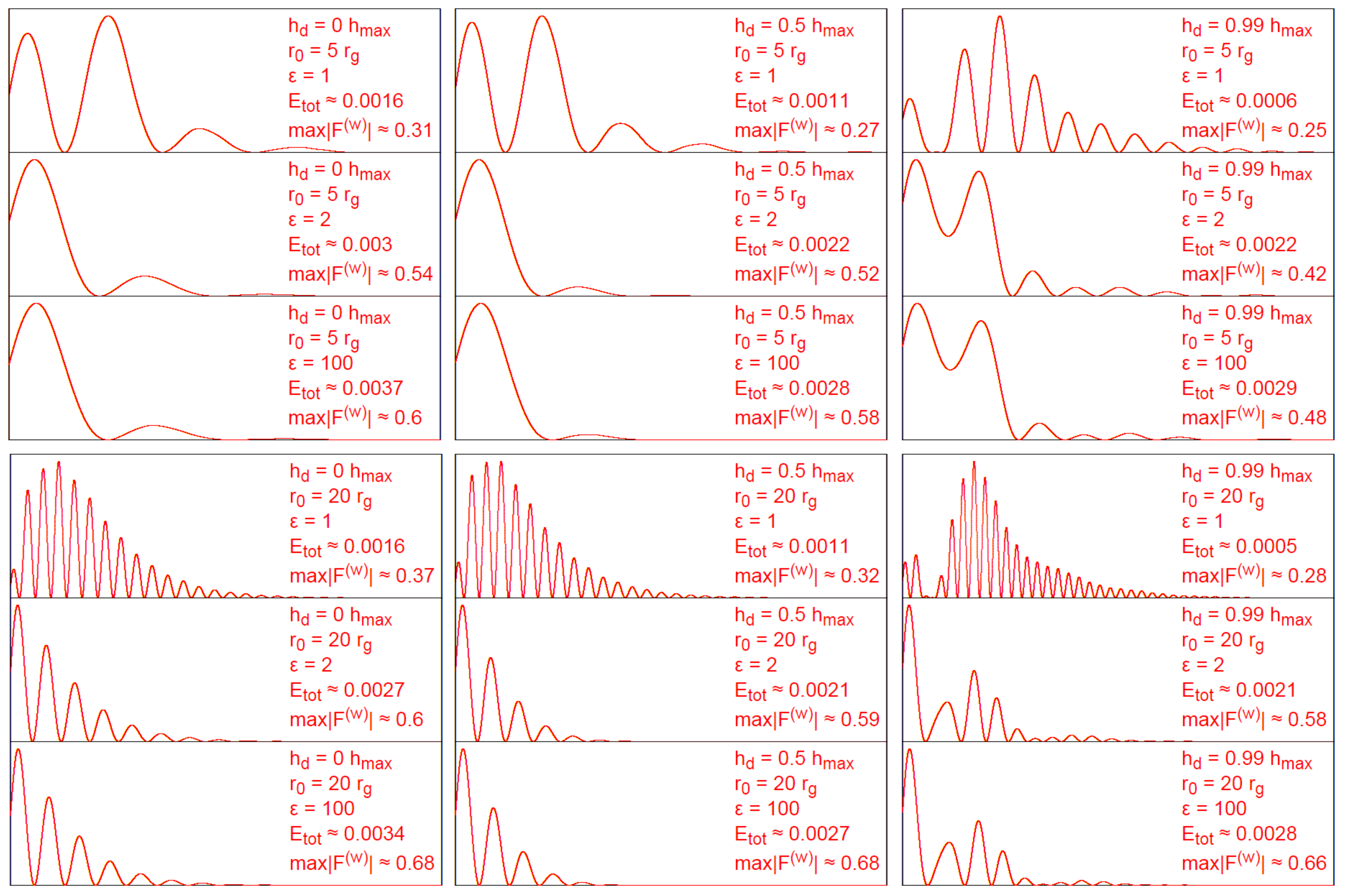}
\caption{The spectral density of the dipole radiation ${\left[F^{({\hat w})}_{0\theta}\right]^2}$.
The parameters $h_d$,  $r_0$, $\epsilon$, total radiated energy ${E_{tot}}$ (in units ${d_0^2/(r_g R^2)}$) and the amplitudes
${|F^{({\hat w})}_{0\theta}|}$ (in units ${d_0/R}$) for all these cases displayed on the panels.
The left-hand sides of panels correspond ${\hat w = 0}$, right-hand sides -- ${\hat w = c/r_g}$.}
\label{R1}\end{figure}

If we know the value of $ {h_ {f}} $ for the photon, we can find the time of the photon propagation
${\Delta t_{f}}$ from the current position of the particle to the observer -- see~(\ref{dt_ph}):
\begin{eqnarray}
\Delta t_{f} = \int\limits_{\xi_{ob}}^{\xi_d} \frac{\sqrt{1-\xi_{ob}}\,\, d\xi}{\xi^2(1-\xi)\sqrt{1-{h_{f}}^2\xi^2(1-\xi)}} .
\label{xi1} \end{eqnarray}
At ${\xi_{ob}\to 0}$ we have: ${\Delta t_{f}\to\infty}$ --
since the time of propagation for the photon to an infinitely for the distant observer tends to infinity.
Therefore, the value ${\xi_{ob}}$ must be greater than zero.

\section{Characteristic features of the spectrum}
\label{s6}

We will explain the characteristic features of the spectrum of the falling into a black hole dipole 
(for low frequencies ${w\le c/r_g}$).
For this purpose, we according to the most general considerations:

1. At the initial moment a distant observer can detect the electric dipole field, as dipole is
sufficiently far away from the black hole.

2. At the final moment (in the limit ${t_{ob}\to\infty}$) the dipole disappears below the horizon of the black hole, all electromagnetic fields
associated with the dipole must be radiated, so a distant observer can not register a dipole field.

The law of decrease of the field for a distant observer was calculated in the previous section.
In general, this law can be approximated by any decaying function.
In the simplest case, we can use the step function:
${F_{0\gamma} = F_{step}\cdot\Theta (t_1 - t_{ob})}$, according to the formula (\ref{F_w_ti_2}) we have:
\begin{eqnarray}
F_{0\gamma}^{(w)} = F_{step}\cdot\frac{1-\cos(wt_1)+\sin(wt_1)}{2w} .
\label{spectr_step} \end{eqnarray}

If we use for this exponent: ${F_{0\gamma} = F_{exp}\cdot\exp(-\alpha t_{ob})\cdot\Theta (t_{ob} - t_1)}$, ${\alpha > 0}$, we have:
\begin{eqnarray}
F_{0\gamma}^{(w)} = F_{exp}\cdot\exp(-\alpha t_1)\cdot\frac{\alpha\sin(wt_1) + w\cos(wt_1) + \alpha\cos(wt_1) - w\sin(wt_1)}{2(\alpha^2 + w^2)} .
\label{spectr_exp} \end{eqnarray}
From the last two formulas it becomes clear why the oscillations in the spectra on Fig.~\ref{R1} and their asymptotic behavior: ${\propto 1/w}$.

\section{Asymptotics of the spectrum of the radiated field at ${{\rm\omega}\to\infty}$}
\label{s7}

To calculate the required asymptotics (at ${w\to\infty}$ -- in dimensionless units) we proceed similarly to our work~\cite{Shatskiy2014}.
The integral in (\ref{F_w_ti_2-3}) splits into two parts:
the first part -- from zero to the observer's time $t_1$ and the second -- from $t_1$ to infinity
(which corresponds to achievement by the particle the horizon). 
Moreover, the time $t_1$ is chosen so that the corresponding radial coordinate of the particle $r_1$ differs from $r_g$ by a small amount
${\delta_1:= r_1-r_g << r_g}$.

The potential ${A_\gamma}$ in the second part of the integral, according to (\ref{inv_A_d1}), is equal to ${\sim (d_0/R) \sqrt{\delta_1/r_g}}$.
In addition, according to (\ref{dt_ob}), the element of time $dt_{ob}$ is equal to ${\sim r_g\, dr/\delta_1}$.
Then in the main approximation by small value $\delta_1$ the second part of the integral in (\ref{F_w_ti_2-3})
will be small compared to the first part ${(\propto \sqrt{\delta_1})}$.

\begin{figure*}
\includegraphics[width=0.95\textwidth]{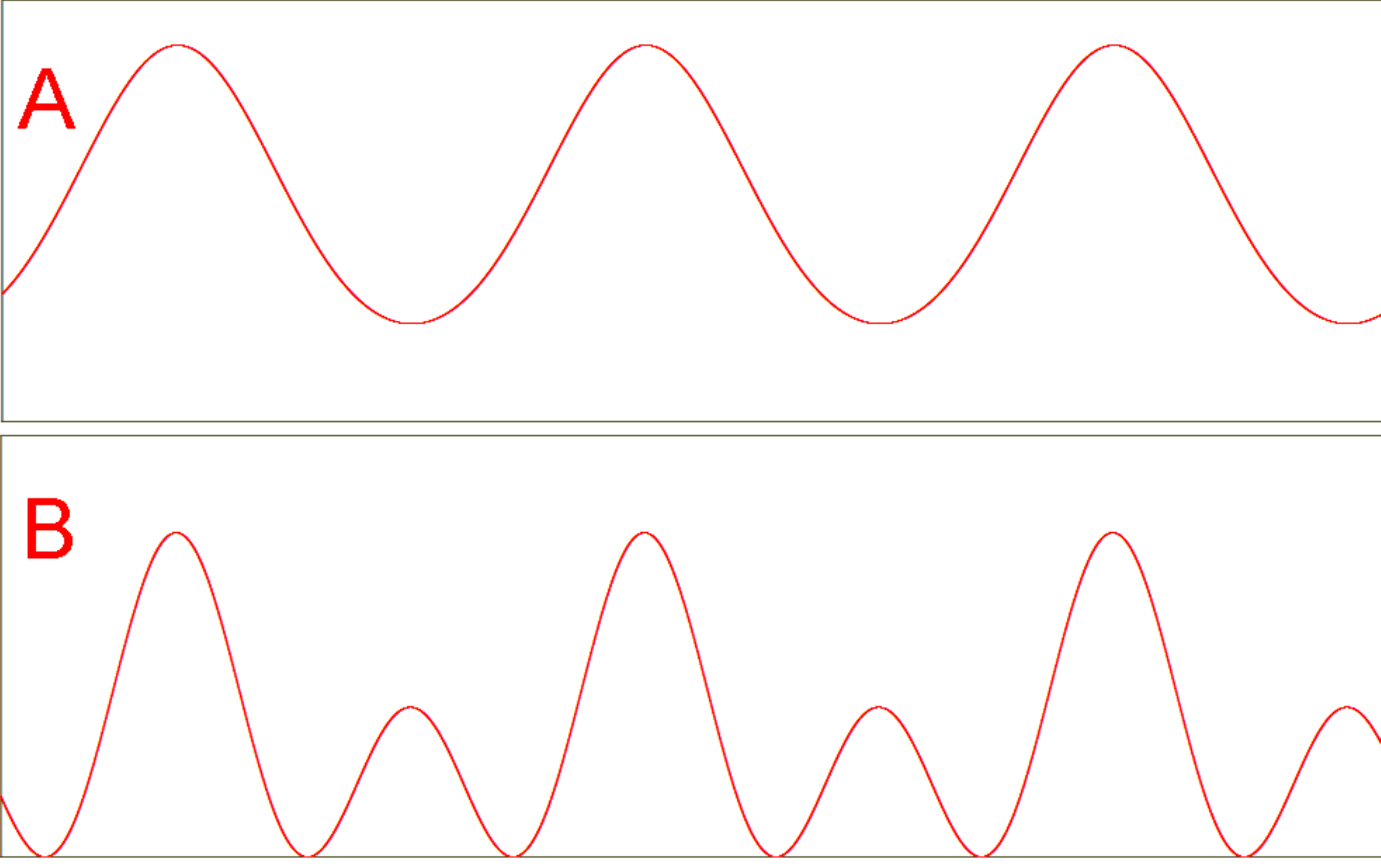}
\caption{The characteristic shape of the spectrum asymptotics at ${w\to\infty}$
(this expression is (\ref{6-4}), which squared and multiplied by $w^2$).
Panels A and B correspond to the different possible relationships between the coefficients
${F_{0\alpha}(0)}$ and ${\sqrt{2}\sum\limits_{n=0}^{\infty} \frac{F_{0\alpha}^{(n)}(0)\, t_1^n}{n!}}$ in (\ref{6-4}).}
\label{R2}\end{figure*}

Since the main contribution to the integrals (\ref{F_w_ti_2}), (\ref{F_w_ti_2-2}) and (\ref{F_w_ti_2-3}) gives the first part
(from $0$ to $t_1$), then the second part we can neglect and consider the first integral in (\ref{F_w_ti_2}) in more detail.
We decompose the integrand function ${F_{0\alpha}(t_ {ob})}$, in a Taylor series with respect to dimensionless time ${t}$:
\begin{eqnarray}
F_{0\alpha}(t) = \sum\limits_{n=0}^{\infty} \frac{F_{0\alpha}^{(n)}(0)\cdot t^n}{n!} .
\label{6-1}\end{eqnarray}
Use the recurrence relation:
\begin{eqnarray}
\int\limits t^n\left[\sin(wt)+\cos(wt)\right] dt = \frac{t^n}{w}\left[\sin(wt)-\cos(wt)\right] -
\frac{n}{w} \int\limits t^{n-1}\left[\sin(wt)-\cos(wt)\right] dt = \quad \nonumber \\
= \frac{t^n}{w}\left[\sin(wt)-\cos(wt)\right] - \frac{n t^{n-1}}{w^2}\, \left[\sin(wt)+\cos(wt)\right] +
\frac{n(n-1)}{w^2} \int\limits t^{n-2}\left[\sin(wt)+\cos(wt)\right] dt = ... \nonumber
\end{eqnarray}
And we use its mathematical limit (the leading order) at ${w\to\infty}$:
\begin{eqnarray}
\int\limits_0^{t_1} t^n\left[\sin(wt)+\cos(wt)\right]\, dt \approx
\frac{t_1^n \left[\sin(wt_1)-\cos(wt_1) +\delta^n_0 \right]}{w} .
\label{6-2}\end{eqnarray}
Using this expression in the decomposition (\ref{6-1}), for (\ref{F_w_ti_2}) we obtain the asymptotic behavior of the spectrum of the radiated field:
\begin{eqnarray}
\lim_{w\to\infty} F_{0\alpha}^{(w)} \approx \frac{\sin(wt_1)-\cos(wt_1)}{2w}
\left(\sum\limits_{n=0}^{\infty} \frac{F_{0\alpha}^{(n)}(0)\, t_1^n}{n!}\right) + \frac{F_{0\alpha}(0)}{2w} . 
\label{6-3}\end{eqnarray}
This shows that the spectral energy density of the attractor ${{\cal E}_w}$ of the radiated field (which proportional to the square of this expression)
is inversely proportional to the square of the frequency: ${{\cal E}_w\propto 1/w^2}$.

The expression (\ref{6-3}) can be rewritten as:
\begin{eqnarray}
\lim_{w\to\infty} F_{0\alpha}^{(w)} \approx \frac{F_{0\alpha}(0)}{2w} -
\frac{\sqrt{2}}{2w}\left(\sum\limits_{n=0}^{\infty} \frac{F_{0\alpha}^{(n)}(0)\, t_1^n}{n!}\right) \cos\left( w t_1 + \frac{\pi}{4} \right) .
\label{6-4}\end{eqnarray}
At ${w\to\infty}$ the shape of this spectrum is almost independent on magnitude of the impact parameter $h_d$ -- see Fig.~\ref{R2}.

\begin{figure*}
\includegraphics[width=0.95\textwidth]{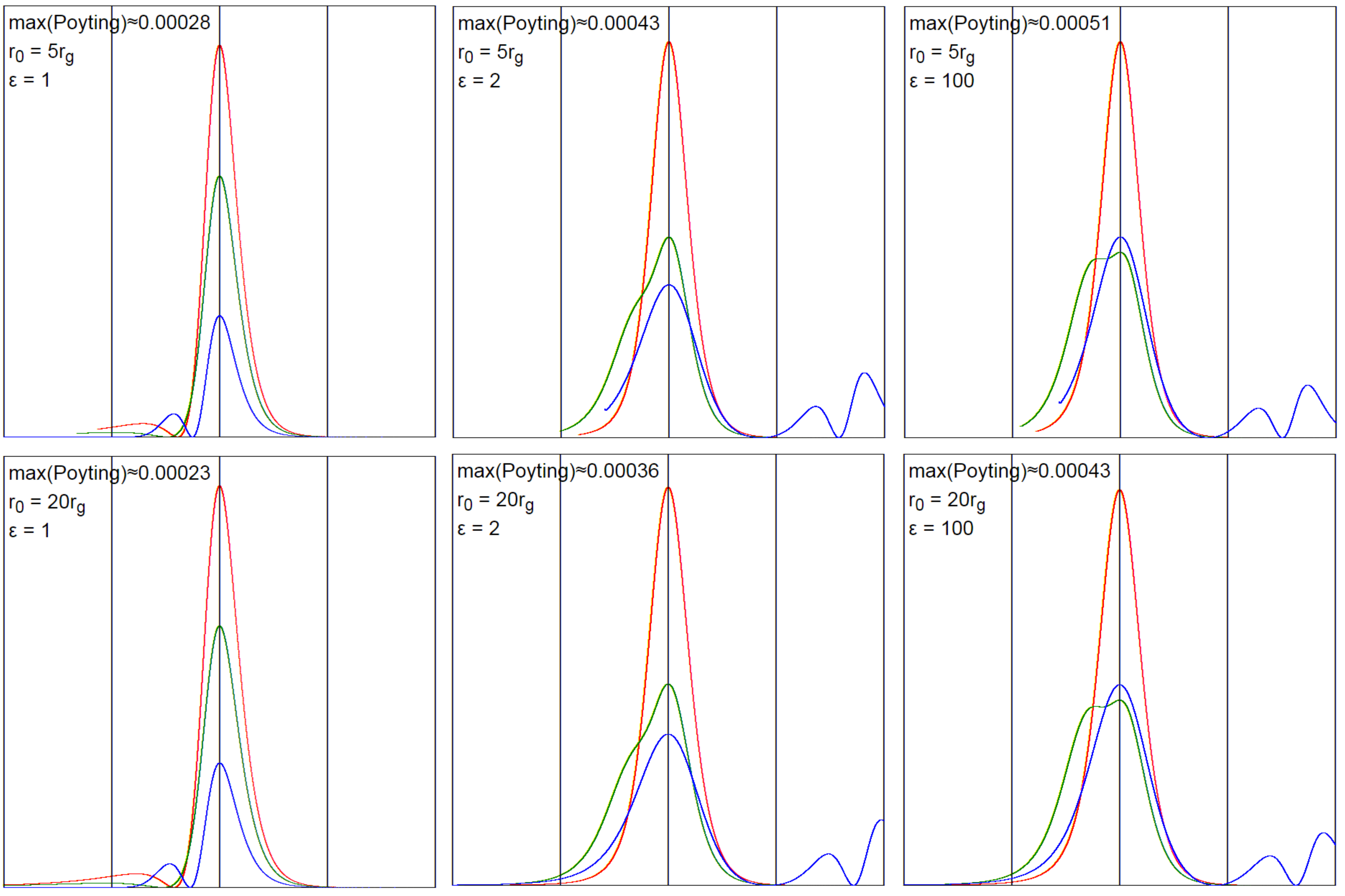}
\caption{Dependencies of the Poynting vectors ${P := F_{0\theta}^2/(4\pi r^2)}$ on ${t_{ob}}$.
The maxima of the curves are placed in the center of the panel, the width of the panels -- ${60 r_g/c}$,
the distance between the vertical lines is equal to ${15 r_g/c}$,
therefore the width of the "bells"\, at half height ${\sim 10 r_g/c}$.
The amplitude maximum (at different $h_d$) indicated in units ${d_0^2/(r^2 R^2)}$ (at ${r_g=1}$ and ${c=1}$).
For each panel, the red curve (with highest amplitude) corresponds to ${h_d = 0}$, the green curve
 -- ${h_d = 0.5h_{max}}$, and the blue curve (with more than one "bell") -- ${h_d = 0.99h_{max}}$.}
\label{R3}\end{figure*}

\section{The dependence of the Poynting vector on time}
\label{s8}

Poynting vector $P$ is given by\footnote{Strictly speaking the value of $P$ is not the Poynting vector,
but only one of its components (at infinity).}:
\begin{eqnarray}
P := \frac{F_{0\theta}^2}{4\pi r^2} , \quad F_{0\theta} = \frac{dA_\theta}{dt_{ob}} .
\label{7-Poyt1}\end{eqnarray}
According to (\ref{dt_ob}), we replace the differentiation with respect ${dt_{ob}}$ to the differentiation with respect $r$ and obtain:
\begin{eqnarray}
\left|F_{0\theta}\right| = \frac{d_0
\sqrt{(1-r_g/r)\left[r^3-(r-r_g)h_f^2\right]} \left[r_g r^3-2 r_g
(r-r_g)(r^2/\epsilon^2+h_d^2)+2(r-r_g)^2 h_d^2\right]}{ 2 R r^{5}
\sqrt{1-r_g/r_{ob}}\left[\sqrt{r^3-(r-r_g)h_f^2} +
\sqrt{r^3-(r-r_g)\left(r^2/\epsilon^2 + h_d^2\right)}\right]} .\quad
\label{7-Poyt2}\end{eqnarray} 
As it was shown in~\cite{Shatskiy2014}, and as it is shown in Fig.~\ref{R3}, profile shape dependent on the 
Pointing vector on time ${t_{ob}}$ is practically independent from
the dipole orbital parameters (the initial radius of falling). And
the profile width at half maximum ${\Delta t_{ob}}$ directly
related with the mass of the black hole: ${\Delta t_{ob}\sim
10r_g/c}$. In general case, non-radial dipole falling into
Schwarzschild black hole the profile form ${P(t_{ob})}$ becomes
dependence on ${\epsilon}$ and $r_0$ at ${h_d>0}$ (see
Fig.~\ref{R3}). At ${h_d\sim 1}$ in the function ${P(t_{ob})}$
appears several local maxima, although the maximum amplitude is
reduced and at the same time slightly reduced the total radiated
energy -- see Fig.~\ref{R1}.

\section{Discussion and conclusions}
\label{s9}

It is seen from the results of the numerical calculations (see. Fig.~\ref{R1}),
the main features of the spectrum are also saved for ${h_d>0}$ 
(asymptotic behavior and an increase in the frequency of the oscillations with increasing $r_0$).
In this case, all the spectral dependence (on the value of $h_d$) is detectable only at small ${w<\sim 1}$,
and at the asymptotic behavior ${(w\to\infty)}$ (as shown in Section~\ref{s7}). 
The behavior of the spectra does not vary. 

The dependence of the amplitude and the total radiated energy on the value $h_d$ is weak,
therefore almost all of the results and conclusions obtained by us for the case ${h_d = 0}$, are also applicable to the general case ${h_d>0}$.
We list here all these results (see~\cite{Shatskiy2014}):

1. In the case of supervisory detection of such spectra (or rather their asymptotics)
we have a real possibility a new and independent way to determine the main characteristics of a black hole - its mass.
In addition, probably also we will be able to determine some properties of the magnetized accreating matter.

2. The main problem, would be too weak energy flow
(for measuring on the available radio frequencies) for observation of such process (dipole falling into a black hole).

3. In addition, the falling of the strongly magnetized compact bodies onto black holes are apparently very rare events,
which also represents a serious obstacle for observations.

4. It is also likely that the scan of the entire range of the radio waves on the subject the search of such spectra lead to the discovery
of new black holes in our galaxy and studying of their properties.

\section*{Acknowledgements}
\label{s10}

Authors are grateful to all workshop participants (in the ASC FIAN and SAI MSU)
for many useful discussions on this topic and valuable suggestions and comments.

This work was supported by RFBR, project code:15-02-00554-a. 

Also work was partially supported within the program of the fundamental
research of Department of Physics of RAS "Interstellar and
intergalactic medium: active and extended objects".



\begin{thebibliography}{15}
\expandafter\ifx\csname natexlab\endcsname\relax\def\natexlab#1{#1}\fi
\expandafter\ifx\csname bibnamefont\endcsname\relax
  \def\bibnamefont#1{#1}\fi
\expandafter\ifx\csname bibfnamefont\endcsname\relax
  \def\bibfnamefont#1{#1}\fi
\expandafter\ifx\csname citenamefont\endcsname\relax
  \def\citenamefont#1{#1}\fi
\expandafter\ifx\csname url\endcsname\relax
  \def\url#1{\texttt{#1}}\fi
\expandafter\ifx\csname urlprefix\endcsname\relax\def\urlprefix{URL }\fi
\providecommand{\bibinfo}[2]{#2}
\providecommand{\eprint}[2][]{\url{#2}}

\bibitem[{Tho(1986)}]{Thorn1998}
\emph{\bibinfo{title}{{Black Holes: The Membrane Paradigm}}}
  (\bibinfo{publisher}{{Kip S. Thorne}}, \bibinfo{year}{1986}),
  \bibinfo{edition}{{Douglas A. MacDonald, Richard H. Price}} ed.

\bibitem[{\citenamefont{Frolov and Novikov}(1998)}]{Frolov1998}
\bibinfo{author}{\bibfnamefont{V.~P.} \bibnamefont{Frolov}} \bibnamefont{and}
  \bibinfo{author}{\bibfnamefont{I.~D.} \bibnamefont{Novikov}},
  \emph{\bibinfo{title}{Black Hole Physics. Basic Concepts and New
  Developments}} (\bibinfo{publisher}{Kluver AP}, \bibinfo{year}{1998}).

\bibitem[{\citenamefont{Linet}(1976)}]{Linet1976}
\bibinfo{author}{\bibfnamefont{B.}~\bibnamefont{Linet}}, \bibinfo{journal}{J.
  Phys. A} \textbf{\bibinfo{volume}{9}}, \bibinfo{pages}{1081}
  (\bibinfo{year}{1976}).

\bibitem[{\citenamefont{{F.J. Zerilli}}(1970)}]{Zerilli1970}
\bibinfo{author}{\bibnamefont{{F.J. Zerilli}}}, \bibinfo{journal}{Phys. Rev. D}
  \textbf{\bibinfo{volume}{2}}, \bibinfo{pages}{2141} (\bibinfo{year}{1970}).

\bibitem[{\citenamefont{{M. Davis, R. Ruffini, W.H. Press and R.H.
  Price}}(1971)}]{Davis1971}
\bibinfo{author}{\bibnamefont{{M. Davis, R. Ruffini, W.H. Press and R.H.
  Price}}}, \bibinfo{journal}{Phys. Rev. Lett.} \textbf{\bibinfo{volume}{27}},
  \bibinfo{pages}{1466} (\bibinfo{year}{1971}).

\bibitem[{\citenamefont{{D.K. Ross}}(1971)}]{Ross1971}
\bibinfo{author}{\bibnamefont{{D.K. Ross}}}, \bibinfo{journal}{Astron. Soc.
  Pacific} \textbf{\bibinfo{volume}{83}}, \bibinfo{pages}{633}
  (\bibinfo{year}{1971}).

\bibitem[{\citenamefont{{S.A. Teukolsky}}(1972)}]{Teukolsky1972}
\bibinfo{author}{\bibnamefont{{S.A. Teukolsky}}}, \bibinfo{journal}{Phys. Rev.
  Lett.} \textbf{\bibinfo{volume}{29}}, \bibinfo{pages}{1114}
  (\bibinfo{year}{1972}).

\bibitem[{\citenamefont{{D.G. Yakovlev}}(1975)}]{Yakovlev1975}
\bibinfo{author}{\bibnamefont{{D.G. Yakovlev}}}, \bibinfo{journal}{JETP}
  \textbf{\bibinfo{volume}{41}}, \bibinfo{pages}{179} (\bibinfo{year}{1975}).

\bibitem[{\citenamefont{{I.G. Dymnikova}}(1977)}]{Dymnikova1977}
\bibinfo{author}{\bibnamefont{{I.G. Dymnikova}}},
  \bibinfo{journal}{Astrophysics and Space Science}
  \textbf{\bibinfo{volume}{51}}, \bibinfo{pages}{229} (\bibinfo{year}{1977}).

\bibitem[{\citenamefont{{Karl Martel and Eric Poisson}}(2008)}]{Martel2008}
\bibinfo{author}{\bibnamefont{{Karl Martel and Eric Poisson}}}
  (\bibinfo{year}{2008}), \eprint{ArXiv: gr-qc/0107104}.

\bibitem[{\citenamefont{{Shatskiy A.A., Novikov I.D., Lipatova
  L.N.}}(2013)}]{Shatskiy2013}
\bibinfo{author}{\bibnamefont{{Shatskiy A.A., Novikov I.D., Lipatova L.N.}}},
  \bibinfo{journal}{{Journal of Experimental and Theoretical Physics}}
  \textbf{\bibinfo{volume}{116}}, \bibinfo{pages}{904} (\bibinfo{year}{2013}).

\bibitem[{\citenamefont{{Shatskii A.A., Novikov I.D., Lipatova
  L.N.}}(2014)}]{Shatskiy2014}
\bibinfo{author}{\bibnamefont{{Shatskii A.A., Novikov I.D., Lipatova L.N.}}},
  \bibinfo{journal}{{Astronomy Reports}} \textbf{\bibinfo{volume}{58}},
  \bibinfo{pages}{39} (\bibinfo{year}{2014}).

\bibitem{Nail2014}
\bibinfo{author}{\bibnamefont{{Nail Khusnutdinov}}},
  \bibinfo{journal}{{PHYSICAL REVIEW}} \textbf{\bibinfo{volume}{D 89}},
  \bibinfo{pages}{024012} (\bibinfo{year}{2014}).

\bibitem[{\citenamefont{{G.C. Graves, D.R. Brill}}(1960)}]{Brill1960}
\bibinfo{author}{\bibnamefont{{G.C. Graves, D.R. Brill}}},
  \bibinfo{journal}{Phys. Rev.} \textbf{\bibinfo{volume}{120}},
  \bibinfo{pages}{1507} (\bibinfo{year}{1960}).

\bibitem[{\citenamefont{{L.D. Landau and E.M. Lifshitz}}(1988)}]{Landau1988}
\bibinfo{author}{\bibnamefont{{L.D. Landau and E.M. Lifshitz}}},
  \emph{\bibinfo{title}{{Course of Theoretical Physics, Volume 2: The Classical
  Theory of Fields}}} (\bibinfo{publisher}{{Nauka, Moscow, 1988;
  Butterworth–Heinemann, Oxford, 1990}}, \bibinfo{year}{1988}).

\bibitem[{\citenamefont{{Ya.B. Zel’dovich and I.D.
  Novikov}}(1967)}]{Novikov1967}
\bibinfo{author}{\bibnamefont{{Ya.B. Zel’dovich and I.D. Novikov}}},
  \emph{\bibinfo{title}{{Relativistic Astrophysics}}}
  (\bibinfo{publisher}{{Nauka, Moscow [in Russian]}}, \bibinfo{year}{1967}).

\end{thebibliography}
\end{document}